\documentclass[12pt]{article}
\usepackage[dvips]{epsfig}
\usepackage{amssymb}

\newcommand{\dual}{\mbox{}^{\ast}}
\newcommand{\dd}{\mbox{\rm d}}
\newcommand{\dD}{{\cal D}}
\newcommand{\Z}{{\mathbb Z}}
\newcommand{\beq}{\begin{equation}}
\newcommand{\eeq}{\end{equation}}
\newcommand{\beqn}{\begin{eqnarray}}
\newcommand{\eeqn}{\end{eqnarray}}
\newcommand{\eq}[1]{(\ref{#1})}
\newcommand{\cD}{{\cal D}}

\newcommand{\cZ}{{\cal Z}}

\newcommand{\itep}
{~\vspace{-1.5cm}
\begin{flushright}
{\large  ITEP-LAT/2002-19}\\
{\large  KANAZAWA-02-36}\\
{\sl December 2, 2002}
\end{flushright}
\vspace{1.0cm}}

\begin{document}

\thispagestyle{empty}

\baselineskip=14pt

\begin{center}

\itep

{\Large\bf Monopole creation operators \\
\vspace{1mm}
as confinement--deconfinement order parameters}

\vskip 1.0cm {\large
V.~A.~Belavin${}^{a}$, M.~N.~Chernodub${}^{a,b}$ and M.~I.~Polikarpov${}^{a}$}\\

\vspace{.4cm}

{ \it
$^a$ ITEP, B. Cheremushkinskaya 25, Moscow, 117259, Russia

\vspace{0.3cm}

$^b$ Institute for Theoretical Physics, Kanazawa University,\\
Kanazawa 920-1192, Japan
}

\end{center}

\begin{abstract}
We study numerically two versions of the monopole creation
operators proposed by Fr\"ohlich and Marchetti. The disadvantage
of the old version of the monopole creation operator is due to
visibility of the Dirac string entering the definition of the
creation operator in the theories with coexisting electric and
magnetic charges. This problem does not exist for the new creation
operator which is rather complicated. Using the Abelian Higgs
model with a compact gauge field we show that both definitions of
the monopole creation operator can serve as order parameters for
the confinement--deconfinement phase transition. The value of the
monopole condensate for the old version depends on the length of
Dirac string. However, as soon as the length is fixed the old
operator certainly discriminates between the phases with condensed
and non--condensed monopoles.
\end{abstract}

\vskip 1cm

\section{Introduction}

The order parameters are useful tools for investigation of phase
transitions. Up to now there is no good definition of the order
parameter for the temperature phase transition in QCD with
dynamical quarks. The traditional order parameters like the
expectation value of the Polyakov line and the tension of the
chromoelectric string work well in the quenched case (no dynamical
quarks) but they fail to distinguish between the phases of the
full QCD. Indeed, even at zero temperature the string spanned on a
quark and an anti--quark can be broken by sea quarks making both
the Polyakov line and the string tension useless candidates for
the order parameter. Moreover, the value of the quark condensate
$\langle {\bar \Psi_q} \Psi_q \rangle$ is strict order parameter
only for massless dynamical quarks.

Below we discuss the quantities which may serve as the order
parameters for full QCD if the monopole (or, "dual
superconductor") confinement mechanism~\cite{DualSuperconductor}
is valid. In this picture the monopoles defined with the help of
the Abelian projection~\cite{AbelianProjections} are supposed to
be condensed in the confinement phase. The monopole condensate
causes a dual analogue of the Abrikosov vortex to be formed
between quarks and anti--quarks. As a result the quarks and
anti--quarks are confined into colorless states. In the
deconfinement phase the monopoles are not condensed and quarks are
not confined. Thus the natural confinement--deconfinement order
parameter is the value of the monopole condensate, which should be
nonzero in the confinement phase and zero in the deconfinement
phase.

There are two (formal) difficulties in defining of the monopole
condensate. At first, the monopole condensate is defined as an
expectation value of the monopole field. However, the monopoles
are the topological defects in the compact Abelian gauge field
and, as a result, the immediate output of the lattice simulations
can not provide us with values the monopole fields. This
difficulty can be overcame by noticing that the lattice output
comes in a form of information about the monopole trajectories.
Then one can apply a known procedure which allows us to rewrite
the path integral over monopole trajectories as an integral over
the monopole fields, and get the monopole condensate.

The second difficulty is that the expectation value of the ({\em
e.g.}, scalar) field $\phi$ should always be zero regardless
whether this field is condensed or not. The reason is very simple:
the path integral includes the integration over all possible
gauges while the charged field is a gauge dependent quantity.
These two problems were solved in Ref.~\cite{FrMa87}, where the
gauge invariant monopole creation operator for compact QED (cQED)
was explicitly constructed. The numerical calculations in the
lattice cQED~\cite{PoWi} and in the Maximal Abelian projection of
the lattice $SU(2)$ gluodynamics~\cite{ChPoVe} confirm that the
expectation value of this operator is an order parameter for the
confinement--deconfinement phase transition. Other types of
the monopole creation operators were investigated in
Refs.~\cite{DiGi}.

However, it noticed recently in Ref.~\cite{FrMa99} that the ``old''
monopole creation operator~\cite{FrMa87} depends on the shape of
the Dirac string in the models with electrically charged dynamical
fields. Exactly the same situation appears in the Abelian
projection of QCD: the off-diagonal gluons become electrically
charged dynamical fields while the diagonal gluons turns into
compact Abelian gauge fields containing monopole singularities.
Below we figure out whether the dependence on the Dirac string in
the ``old'' version of the monopole creation operator is crucial
for the usage of the expectation value of this operator as an
order parameter. Another (``new'') monopole creation operator was
suggested in Ref.~\cite{FrMa99}. This operator does not depend on
the shape of the Dirac string even in the presence of the
dynamical electric charges. However, the numerical
calculations~\cite{StLesna} show that the construction of this
operator is rather complicated and the simulations are time
consuming.

The structure of the this paper is as follows. We explicitly
describe the ``new'' and the ``old'' monopole creation operators
in Sections~2 and 3, respectively. The results of the numerical
calculations of these creation operators in the compact Abelian
Higgs model are presented in Section~4. We demonstrate that both
these operators can be used to detect confinement--deconfinement
phase transitions in the theories with charged matter. Our
conclusions are presented in Section~5.

\section{``Old'' monopole creation operator.}

The gauge invariant creation operator $\Phi$ was suggested by
Dirac \cite{Dirac}:
\beqn
\label{Dirac}
\Phi = \phi (x) \exp\left\{i \int E_k(\vec{x}-\vec{y}) A_k(\vec{y}) \,
d^3 y\right\} \,\, ,
\eeqn
here $\phi(x)$ and $A_k (x)$ are the electrically charged field
and the gauge potential, which transform under the gauge
transformations as
\beqn
\label{GT}
\phi (x) \to \phi (x) e^{i\alpha(x)}\,,\qquad
A_k(x) \to A_k(x) + \partial_k \alpha(x)\,.
\eeqn

The Coulomb field, $E_k(x)$, satisfies the equation:
\beqn
\partial_k E_k = \delta^{(3)} (x)\, .
\eeqn
It is easy to see that the operator $\Phi$, eq.~\eq{Dirac}, is invariant under
the gauge transformations \eq{GT}.

The Fr\"ohlich--Marchetti construction \cite{FrMa87} of the monopole
creation operator in cQED is based on eq.~\eq{Dirac}. At first
step the partition function of cQED is transformed to a dual
representation. For the general form of the cQED action it can be
shown \cite{ChPoVe} that the dual theory is an Abelian Higgs model
(AHM) in the limit when both the Higgs boson mass and the gauge boson
mass are infinite. In this theory the Higgs field, $\phi_x$,
corresponds to the monopole defect in the original cQED. The gauge field
$\dual B$ is dual to the original gauge field $\theta$. Thus the gauge
invariant creation operator \eq{Dirac} for the AHM model
corresponds to the monopole creation operator in the original
cQED. The explicit expression for this operator on the lattice is
({$cf$. eq.\eq{GT}}):
\beq
\Phi^{\mathrm{mon,old}}_x = \phi_x \, e^{i (\dual B, \dual H_x)}\,,
\label{Phi}
\eeq
where $\dual H_x$ is the Coulomb field of
the monopole, $\delta \dual H_x = \dual \delta_x$, and
$\dual \delta_x$ is the discrete $\delta$--function defined on the
dual lattice. Here and below we will use the differential
form notations on the lattice: $(a,b) = \sum_c a_c b_c$  is the
scalar product of the forms $a$ and $b$ defined on the $c$--sells;
$(a,a) \equiv {||a||}^2$ is the norm of the form $a$; $d$
is the forward derivative (an analog of the gradient);
$\delta$ is the backward derivative (an analog of the divergence)
and $\dual$--operation transfers a form to the dual lattice.
For a description of the language of the differential forms on
the lattice see, $e.g.$, review~\cite{Review}.

Performing the inverse duality transformation for the expectation
value of the creation operator \eq{Phi} we get
the expectation value, $\langle\Phi^{mon}\rangle$, of this operator
in cQED:
\beqn
\langle \Phi^{\mathrm{mon,old}} \rangle \!
=
\! {1 \over \cZ}
\int\nolimits_{- \pi}^{\pi} \!\!\! \dD \theta \, \exp\{ -S(\dd \theta + W)\}\,,
\qquad
\cZ \!
=
\! \int\nolimits_{- \pi}^{ \pi} \!\!\! \cD \theta \, \exp \{-
S(d\theta) \}\,, \label{original}
\eeqn
where $\dd\theta$ is the plaquette angle, and $S$ is the periodic lattice
action, $S(\dd\theta+2\pi n)=S(\dd\theta)$, $n \in \Z$. The form  $W=2
\pi \delta \Delta^{-1}\dual (H_x - \omega_x)$ depends on the
Dirac string $\dual \omega_x$ defined on
the dual lattice. The Dirac strings start and end on the monopoles and
anti--monopoles, $\delta \dual \omega_x = \dual \delta_x$. The
operator $\Delta=\dd \delta+\delta \dd$ is the lattice Laplacian. The numerical
investigation of this creation operator in cQED shows \cite{PoWi} that
it can be used as the confinement--deconfinement order parameter.

The operator \eq{Phi} is well defined for the theories without
dynamical matter fields. However, if an electrically charged
matter is added, then the creation operator \eq{Phi} depends on
the position of the Dirac string. To see this fact
let us consider the compact Abelian Higgs model with
the Villain form of the action:
\beqn
\cZ_{AHM} = \int^\pi_{-\pi} \!\!\!\dD \theta \int^\pi_{-\pi} \!\!\!\dD  \varphi
\sum_{n \in \Z(c_2)} \sum_{l \in \Z(c_1)} \,
\exp\Bigl\{- \beta ||{\mathrm d} \theta + 2\pi n||^2
- \gamma || {\mathrm d} \varphi + q \theta + 2\pi l||^2\Bigr\}\,.
\label{ZAHM}
\eeqn
Here $\theta$ is the compact Abelian gauge field and $\varphi$ is
the phase of the dynamical Higgs field. The integer $q$ is the charge of
the Higgs field. For the sake of simplicity we consider the London
limit (the Higgs mass is infinitely large while the Higgs condensate
is fixed).

Let us perform the Berezinsky-Kosterlitz-Thouless (BKT)
transformation~\cite{BKT} with respect to the compact gauge field
$\theta$:
\beqn
\dd \theta + 2 \pi n = \dd A + 2 \pi \delta \Delta^{-1} j\,, \quad
\mbox{with} \quad
A = \theta + 2 \pi \delta \Delta^{-1} m[j] + 2 \pi k\,.
\label{BKTtheta}
\eeqn
Here $A$ is the non--compact gauge field, $\dual m[j]$ is a surface on the
dual lattice spanned on the monopole current $\dual j$ ($\delta
\dual m[j] = \dual j$) and $k$ is the integer--valued vector
form\footnote{A detailed description
of the duality and BKT transformations in terms of the
differential forms on the lattice can be found, $e.g.$, in
Ref.~\cite{Review}.}. We substitute eqs.\eq{BKTtheta} in
eq.\eq{ZAHM} and shift of the integer variable, $l \to l + q k$.

Next we perform the BKT transformation with respect to the
compact scalar field $\varphi$:
\beqn
\dd \varphi + 2 \pi l = \dd \vartheta + 2 \pi \delta \Delta^{-1} \sigma\,,
\quad \mbox{with} \quad
\vartheta = \varphi + 2 \pi \delta \Delta^{-1} s[\sigma] + 2 \pi p\,.
\label{BKTphi}
\eeqn
Here $\vartheta$ is the non--compact scalar field, $\dual s[\sigma]$ is
a $3D$ hyper--surface on the dual lattice spanned on the closed surface
$\dual \sigma$ ($\delta \dual s[\sigma] = \dual \sigma$) and $p$
is the integer--valued scalar form.

Substituting eqs.~(\ref{BKTtheta},\ref{BKTphi}) into the partition
function~\eq{ZAHM} and integrating the fields $A$ and $\varphi$
we get the representation of the compact AHM in terms of the monopole
currents $\dual j$ and world sheets of Abrikosov strings $\dual \sigma$
("the BKT--representation"):
\beqn
\cZ_{AHM} \propto \cZ_{BKT} = \!\!\!\!
\sum_{\stackrel{\dual j \in \Z(\dual c_3)}{\delta \dual j = 0}}
\sum_{\stackrel{\dual \sigma_j \in \Z(\dual c_2)}{\delta \dual \sigma_j =q \dual j}}
\!\!\!\!
\exp\Bigl\{\!\!\!\! & - & \!\!\!\! 4 \pi^2 \beta \,
\Bigl(j,
\frac{1}{\Delta + m^2}
j\Bigr)
\label{ZAHM2}
-
4 \pi^2 \gamma \,
\Bigl(\sigma_j,
\frac{1}{\Delta + m^2}
%
\sigma_j\Bigr)
\Bigr\}\,,
\eeqn
the dual surface variable
$\dual \sigma_j = \dual \sigma + q \dual m[j]$ is spanned
$q$--times on the monopole current $j$,
$\delta \dual \sigma_j =q \dual j$, since the flux of the
magnetic monopole having an unit magnetic charge can be taken out by
$q$ strings carrying the elementary flux $2 \pi \slash q$.
The mass of the gauge boson $\theta$ is
$m = q \sqrt{\gamma \slash \beta}$.

The BKT--representation \eq{ZAHM2} of the AHM partition function
\eq{ZAHM} can also be transformed into the dual representation using
simple Gaussian integrations. We use two dual compact fields $\dual B$
(vector field) and $\dual \xi$ (scalar field) in order to represent the
closeness properties of the currents $\dual \sigma_j$ and $\dual
j$, respectively. We also introduce two dual non--compact fields, $\dual F$
(vector field) and $\dual G$ (rank-2 tensor field), in order to get
a linear dependence on, correspondingly, the currents $\dual \sigma_j$ and
$\dual j$ under the exponential function:
\beqn
\cZ_{BKT} & = & {\mathrm {const.}} \,
\int^\infty_{-\infty} \dD \dual F \, \int^\infty_{-\infty} \dD \dual G
\int^\pi_{-\pi} \dD \dual B \, \int^\pi_{-\pi} \dD \dual \xi
\sum_{\dual j \in \Z(\dual c_3)} \sum_{\dual \sigma_j \in \Z(\dual c_2)}
\nonumber \\
& & \exp\Bigl\{ - \dual \beta  \, (\dual G, (\Delta + m^2) \dual G)
            - \dual \gamma \, (\dual F, (\Delta + m^2) \dual F)
\nonumber \\
& & + i (\dual F, \dual \sigma_j) + i (\dual G, \dual j)
+ i (\dual B, \delta \dual \sigma_j - q \dual j)
- i (\dual \xi, \delta \dual j) \Bigr\}\,,
\eeqn
where
\beqn
\dual \beta = \frac{1}{16 \pi^2 \gamma}\,,\quad
\dual \gamma = \frac{1}{16 \pi^2 \beta}\,.
\eeqn
Note that in this representation the integer variables $\dual \sigma_j$
and $\dual j$ are no more restricted by the closeness relations.
Therefore we can use the Poisson summation formula with respect
to these variables and integrate out the fields $\dual F$ and $\dual
G$. Finally, we obtain the dual {\it field} representation of the partition
function \eq{ZAHM}:
\beqn
\cZ_{BKT} & \propto & \cZ_{dual\, field} =
\int^\pi_{-\pi} \dD \dual B \, \int^\pi_{-\pi} \dD \dual \xi
\sum_{\dual u \in \Z(\dual c_3)} \sum_{\dual v \in \Z(\dual c_2)}
\nonumber \\
& & \exp\Bigl\{ - \dual \beta  \, \left(\dd \dual B + 2 \pi \dual u,
(\Delta + m^2) \, (\dd \dual B + 2 \pi \dual u)\right)
\label{ZAHMdual} \\
& & - \dual \gamma \, \left(\dd \dual \xi + q \dual B + 2 \pi \dual v,
(\Delta + m^2) \, (\dd \dual \xi + q \dual B + 2 \pi \dual v)\right)
\Bigr\}\,, \nonumber
\eeqn
where $\dual u$  and $\dual v$ are the integer valued forms
defined on the plaquettes and links of the dual lattice,
respectively. Clearly, this is the dual Abelian Higgs model with
the modified action. The gauge field $\dual B$ is compact and the
radial variable of the Higgs field is frozen. The model is in
the London limit and the dynamical scalar variable is
the phase of the Higgs field $\dual \xi$.

Thus in the presence of the dynamical matter the dual gauge field
$\dual B$ becomes compact\footnote{Another way to establish this fact is
to realize that the pure compact gauge model is dual to the
non--compact $U(1)$ with matter fields (referred above as the
(dual) Abelian Higgs model). Reading this relation backwards one
can conclude that the presence of the matter field leads to the
compactification of the dual gauge field~$\dual B$.}.
The compactness of the dual gauge field implies that it
transforms under the gauge transformations as:
\beq
\dual B \rightarrow \dual B + \dd \dual \alpha + 2 \pi \dual k\,,
\label{cgauge}
\eeq
where the integer valued field $k$ is chosen in such a way that
$\dual B \in (-\pi,\pi]$.

One can easily check that the operator \eq{Phi} is not invariant under
these gauge transformations:
\beq
\Phi^{\mathrm{mon,old}}_x (H) \to \Phi^{\mathrm{mon,old}}_x (H)
\, e^{2 \pi i (\dual k, \dual H_x)}\,.
\label{change}
\eeq

\section{``New'' monopole creation operator.}

The invariance of the operator \eq{Phi} under the gauge
transformations \eq{cgauge} can be achieved if and only if
the function $\dual H_x$ is an integer--valued form.
Thus, if we take into account the
Maxwell equation, $\delta \dual H_x = \dual \delta_x$, we find
that $\dual H_x$ should
be a string attached to the monopole ("the Mandelstam string"): $\dual H_x \to
\dual j_x$, $\dual j_x \in \Z$, $\delta \dual j_x = \dual \delta_x$. The string
must belong to the three--dimensional time--slice. However, one can
show~\cite{FrMa99} that for a fixed string position the operator $\Phi$
creates a state with an infinite energy. This difficulty may be
bypassed~\cite{FrMa99}
by summation over all possible positions of the Mandelstam strings with
a measure $\mu(\dual j)$:
\beqn
\Phi^{\mathrm{mon,new}}_x = \phi_x \,
\sum_{\stackrel{j_x \in \Z}{\delta \dual j_x = \dual \delta_x}} \mu(\dual j_x)
\, e^{i (\dual B, \dual j_x)}\,. \label{Phi:new}
\eeqn
The summation over the strings provides an entropy factor which
cancels the energy suppression.
An example of a ``reasonable'' measure $\mu(j_x)$ is~\cite{FrMa99}:
\beqn
\mu(\dual j_x) = \exp\Bigl\{ - \frac{1}{2 \kappa}
{||\dual j_x||}^2\Bigr\}\,.  \label{Good:mu}
\eeqn
If the Higgs field $\phi$ is
$q$--charged ($q \in \Z$), the summation in eq.\eq{Phi:new} should be
taken over $q$ different strings with the unit flux.
Measure \eq{Good:mu} corresponds to the dual formulation of
the $3D$ XY--model with the Villain action:
\beq
S_{XY}(\chi, r) = {\kappa \over 2} || \dd \chi -2\pi B + 2 \pi r||^2
\,.
\label{xy}
\eeq
The Mandelstam strings correspond to the ordinary vortices in the
XY-model~\eq{xy}.

The two point correlation function in the XY-model~\eq{xy} is given by
\beq
\langle e^{i \chi_x} e^{-i \chi_y} \rangle =
\sum_{j_{xy} \in \Z} \mu(\dual j_{xy})\,
e^{i (\dual B, \dual j_{xy})}
\,,
\label{xy:corr}
\eeq
where $\delta \dual j_{xy} = \dual \delta_x - \dual \delta_y$.
For sufficiently large coupling $\kappa$ and sufficiently small field $B$
we get
\beq
\langle e^{i \chi_x} e^{-i \chi_y} \rangle \rightarrow
\mbox{const}\,,
\eeq
as $|x-y| \rightarrow \infty$. So that the correlation function \eq{xy:corr}
might yield an appropriate measure of the Mandelstam strings.

Performing the inverse duality transformation of the creation
operator~\eq{Phi:new} in the dual representation of the compact
AHM~\eq{ZAHMdual} we get the expectation value
of this operator in terms of the compact gauge field $\theta$
in the original representation~\eq{ZAHM}:
\beqn
\langle\Phi^{\mathrm{mon,new}} \rangle ={1 \over Z}
\sum_{\stackrel{\dual j \in \Z(\dual c_3)}{\delta \dual j =
\dual \delta_x}}
\int_{- \pi}^{ \pi} D \theta
\exp \Bigl\{ -{1 \over 2 \kappa} ||\dd \dual j||^2
- S(\dd \theta + {2 \pi \over q}  \tilde j) \Bigr\} \,.
\label{new}
\eeqn
The current $\tilde j \equiv \dual{}^{(3)} \dual{}^{(4)} j$ is
defined via a double duality operation which is first applied in
the 3D time slice and then in the full 4D space.

We thus defined ``old'' \eq{original} and ``new'' \eq{new} monopole
creation operators.

\section{Numerical results}

Below we present results of the numerical simulation of the old
and the new monopole creation operators. We study them in the
Abelian Higgs model with compact gauge field and infinitely deep
potential for the Higgs field corresponding to the London limit.
This is the simplest model containing both the monopoles and the
electrically charged fields. We used the Wilson form of the action
which is more suitable for the numerical simulations than the
Villain action used for the analytical calculations above:
\beqn
\cZ_{AHM}^{\mathrm{W}} = \int^\pi_{-\pi} \!\!\!\dD \theta \int^\pi_{-\pi} \!\!\!\dD
\varphi \, e^{\beta \cos {\mathrm d} (\theta)
+ \gamma \cos ({\mathrm d} \varphi + q \theta) }\,.
\label{ZAHM_W}
\eeqn
The expectation values of the old and new monopole creation
operators in the Wilson form are given by
\beqn
\langle\Phi^{\mathrm{mon,old}} \rangle & = &{1 \over Z}
\int_{- \pi}^{ \pi} D \theta
\exp \{ \beta \cos (\dd \theta
+ 2 \pi \delta \Delta^{-1}\dual (H_x - \omega_x))
+ \gamma \cos (q \, \theta)\}\,,
\label{SrPhi:old_w}\\
\langle\Phi^{\mathrm{mon,new}} \rangle & = & {1 \over Z}
\sum_{\stackrel{\dual j \in \Z(\dual c_3)}{\delta \dual j = \dual \delta_x}}
\int_{- \pi}^{ \pi} D \theta
\exp \{ -{1 \over 2 \kappa} ||\dd \dual j||^2
+ \beta \cos \left(\dd \theta + {2 \pi \over q}  \tilde j\right)
+ \gamma \cos (q \, \theta)\}\,, \label{SrPhi:new_w}
\eeqn
respectively. We have fixed the unitary gauge, therefore the phase
of the Higgs field is eaten up by the corresponding gauge
transformation.

The value of the monopole order parameter, $\langle \phi \rangle$,
corresponds to the minimum of the (effective constraint)
potential on the monopole field. This potential can be estimated
as follows:
\beqn
V_{\mathrm{eff}}(\Phi) = - \ln \Bigl[\Bigl\langle \delta\Bigl(\Phi
- \Phi^{\mathrm{mon,new/old}}\Bigr) \Bigr\rangle\Bigr]\,.
\label{eff:potential}
\eeqn

We are studying the model with $q=7$. The compact Abelian Higgs
model with multiple charged Higgs fields is known to have a
nontrivial phase structure containing the Coulomb, Higgs and
Confinement phases~\cite{Fradkin}. In this paper we concentrate on
the phase transition between the confinement and the Coulomb
phases.

We simulated the 4D Abelian Higgs model with anti-periodic
boundary conditions in the $3D$ space (the single monopole charge
can not exist in the finite volume with periodic boundary
conditions). For the new creation operator we perform our
calculations on the $4^4,6^4,8^4$ lattices and the coupling
constant $\gamma = 0.3$. For each configuration of 4D fields we
simulated 3D model to get the ensemble of the Mandelstam strings
with the weight $\mu(j_x)$. We generated 60 statistically
independent 4D field configurations, and for each of these
configurations we generated 40 configurations of 3D Mandelstam
strings.

According to eq.~\eq{xy} the weight function \eq{Good:mu}
corresponds to the 3D XY--model with the Villain action. The XY
model has the phase transition at $\kappa_c(B=0) \approx
0.32$~\cite{Kog77}. Our numerical observations show that in
presence of the external field $B$ the critical coupling constant
gets shifted, $\kappa_c^B \approx 0.42$.

\begin{figure}[!htb]
\begin{center}
\begin{tabular}{cc}
\epsfxsize=5.2cm \epsffile{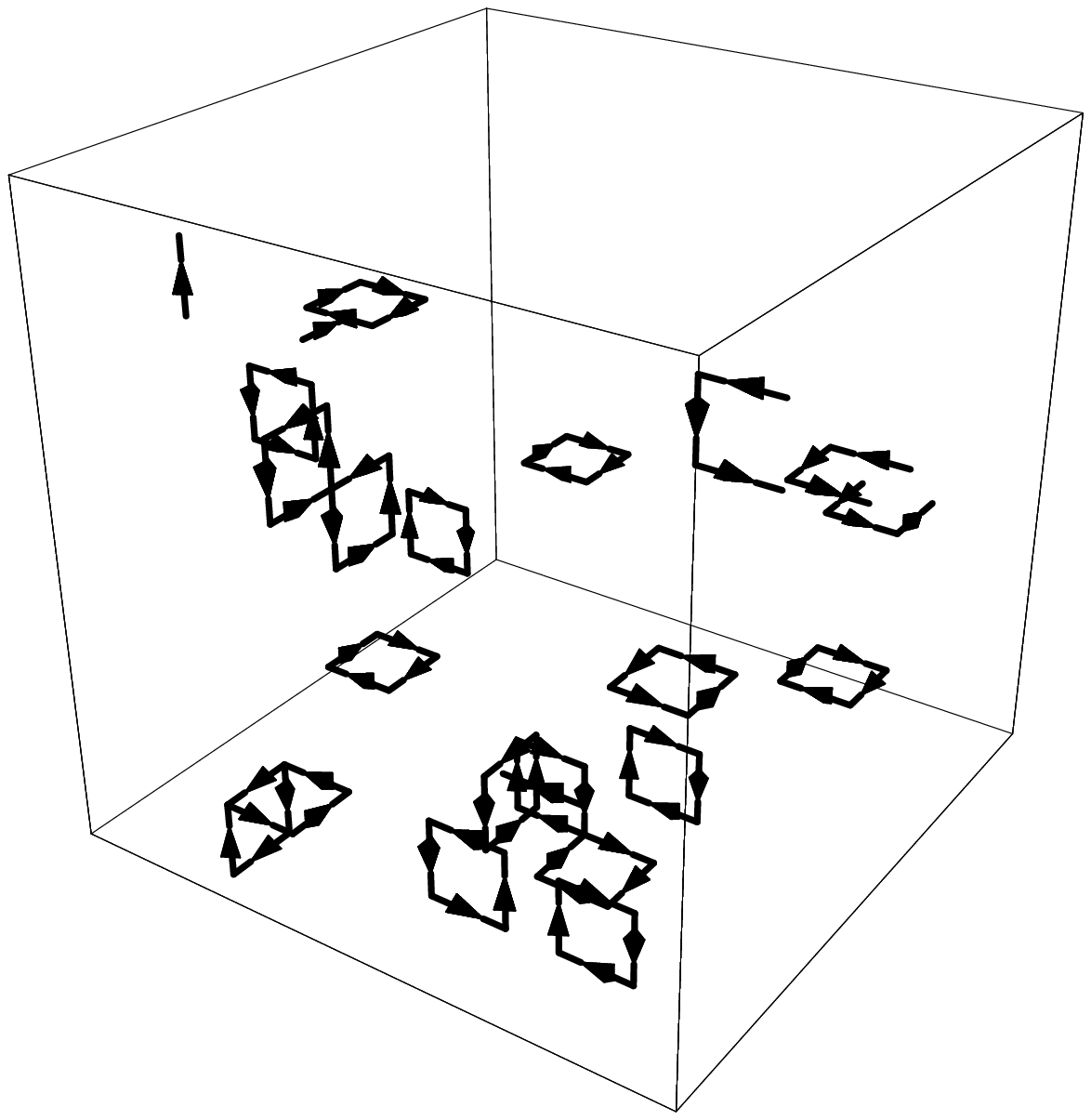} &
\epsfxsize=5.2cm \epsffile{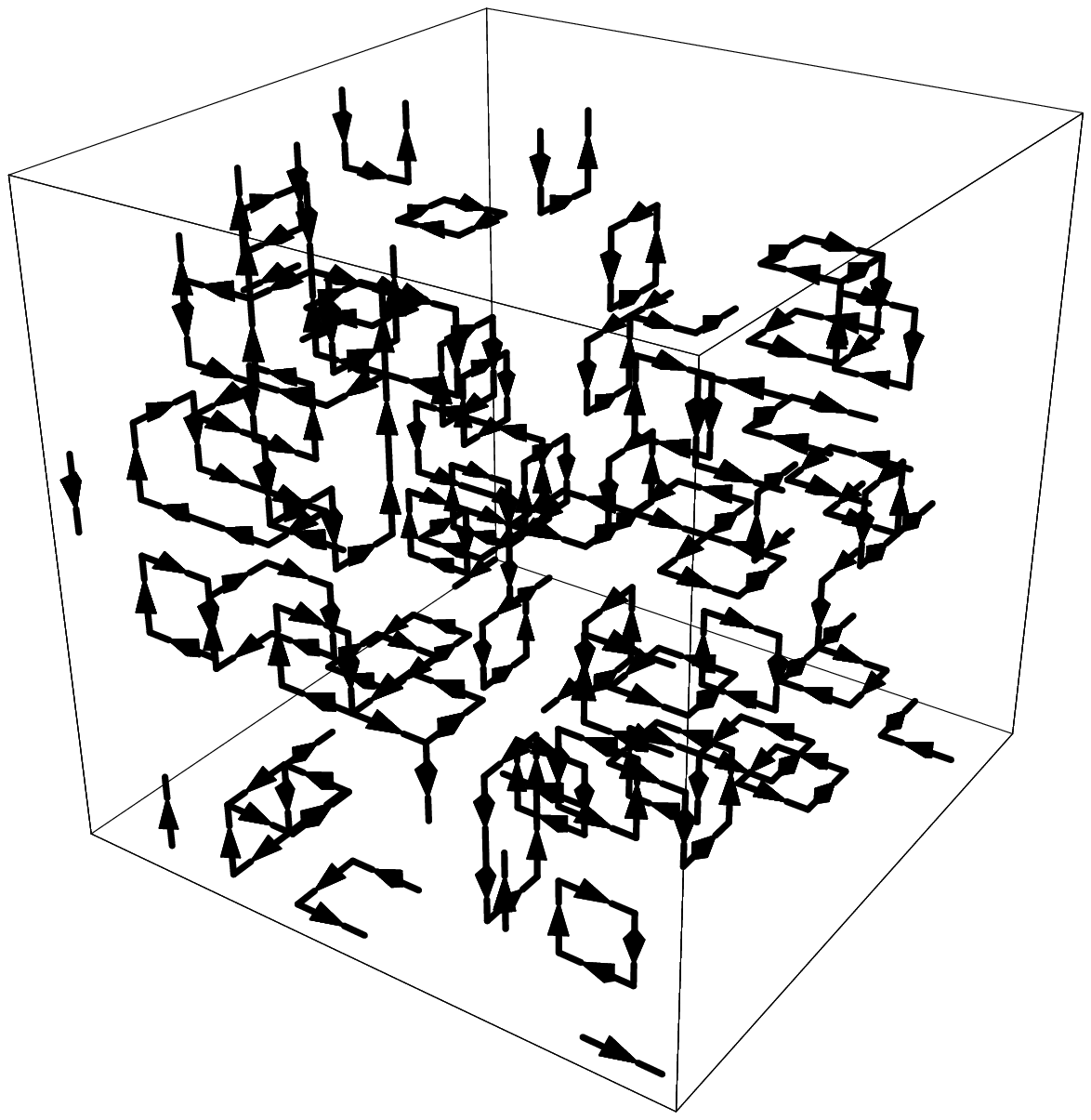} \\
(a) &
(b)
\end{tabular}
\end{center}
\caption{The typical Mandelstam strings in the auxiliary $3D$
model in (a) the ordered phase (no string condensate, $\kappa=0.3$)
and (b) disordered phase (nonzero string condensate, $\kappa=0.5$).}
\label{fig:loops}
\end{figure}

Typical configurations of the Man\-del\-stam strings corresponding
to the phase with the condensate of these strings (disordered
phase, large $\kappa$) and to the phase without the string
condensate (ordered phase, small $\kappa$) are shown in
Figures~\ref{fig:loops}(a,b).

One can expect that the operator~\eq{Phi:new} plays the role of
the order parameter in the phase, where the Mandelstam strings are
condensed ($\kappa>\kappa_c$).
\begin{figure}[!htb]
\begin{center}
\begin{tabular}{cc}
\epsfxsize=6.cm \epsffile{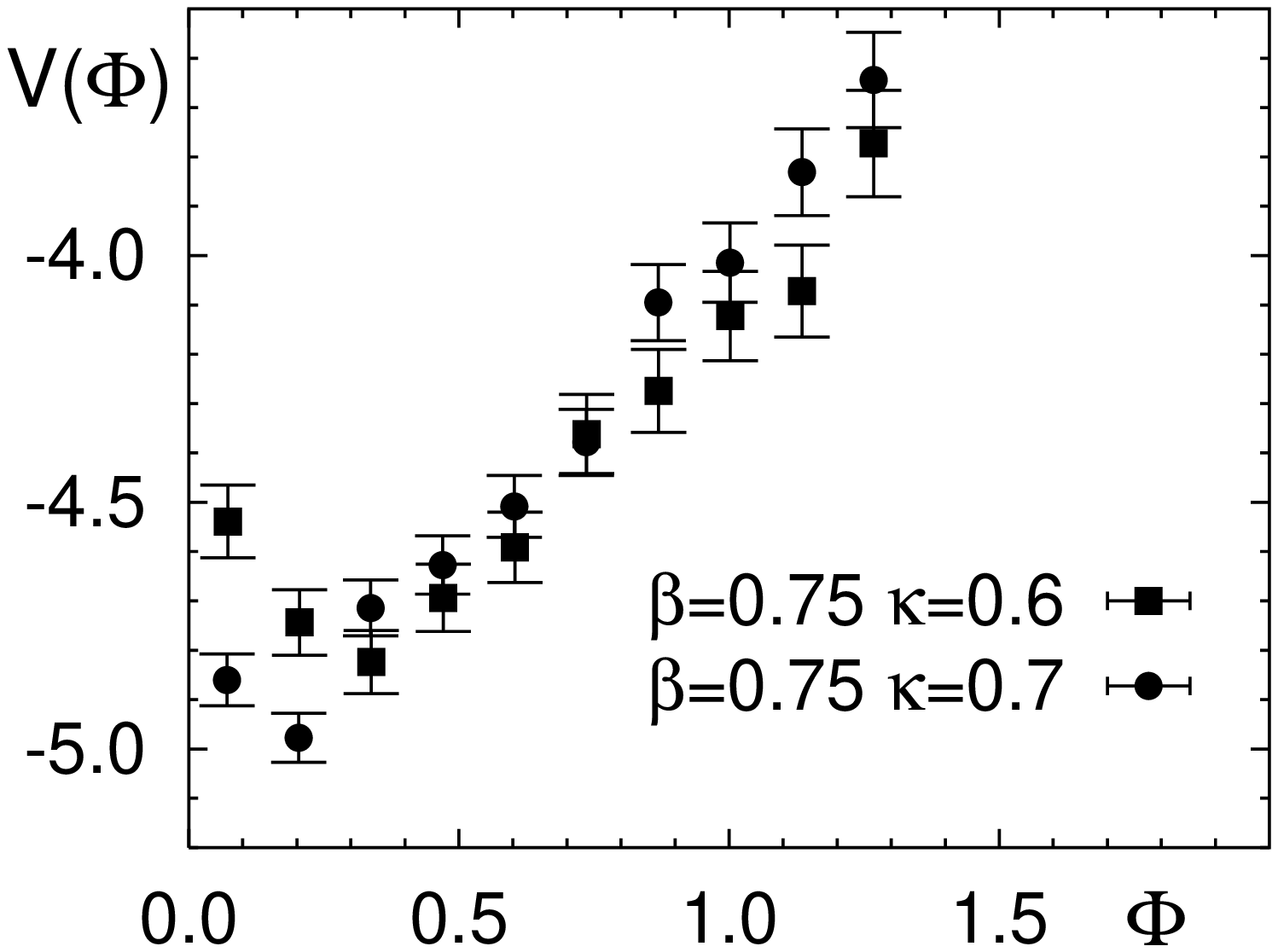} &
\epsfxsize=6.cm \epsffile{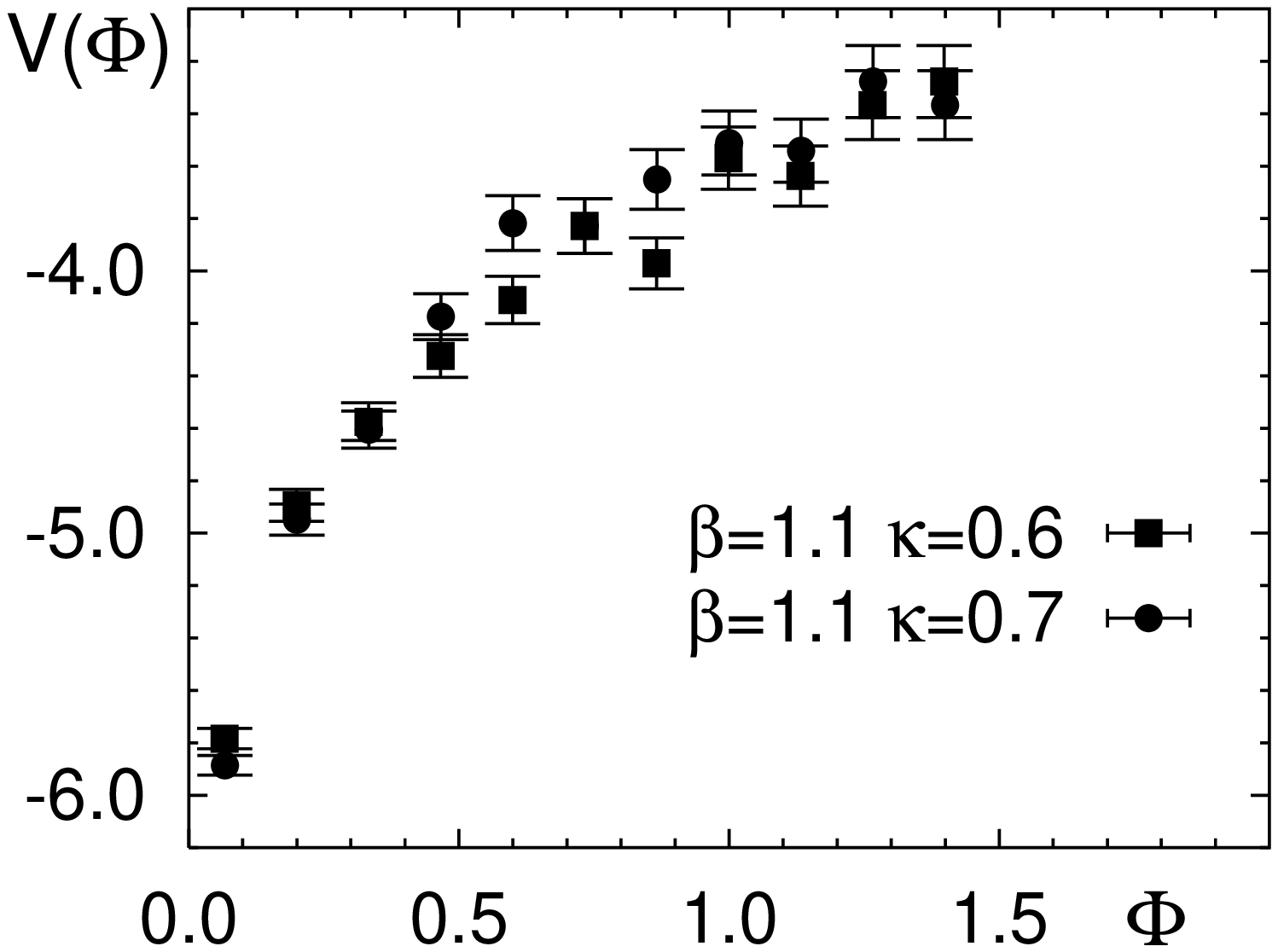} \\
(a) &
(b)
\end{tabular}
\end{center}
\caption{The effective monopole potential~\eq{eff:potential}
in (a) confinement and (b) deconfinement phases.}
\label{fig:potentials}
\end{figure}
In Figures~\ref{fig:potentials} we
present the effective potential \eq{eff:potential} in the confinement
($\beta=0.85$) and deconfinement ($\beta=1.05$) phases. The potential is
shown for two values of the $3D$ coupling constants $\kappa > \kappa_c$
corresponding to high densities of the Mandelstam strings. In the
confinement phase, Figure~\ref{fig:potentials}(a), the potential
$V(\Phi)$ has a Higgs form signaling the monopole
condensation. According to our numerical observations this statement
does not depend on the lattice volume. In the deconfinement phase,
Figure~\ref{fig:potentials}(b), the potential has minimum at $\Phi = 0$
which indicates the absence of the monopole condensate.

\begin{figure}[!htb]
\vspace{2mm}
\begin{center}
\epsfxsize=8.0cm \epsffile{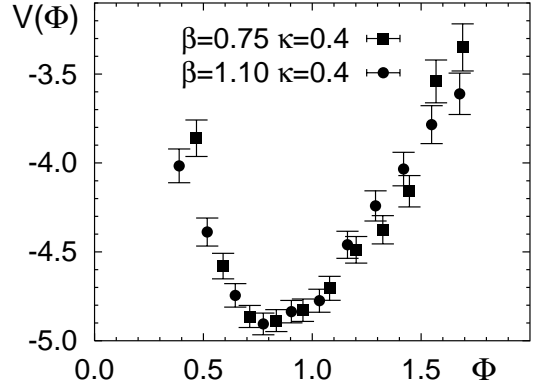}
\end{center}
\caption{The effective monopole potential~\eq{eff:potential}
in the low--$\kappa$ region of the $3D$ model.}
\label{fig:smallk}
\end{figure}
For small values of the $3D$ coupling constant $\kappa$ (in the
phase where Mandelstam strings $j_x$ are not condensed), we found
(Figure~\ref{fig:smallk}) that the potential $V(\Phi)$ has the
same behavior for the both phases of $4D$ model. Thus the operator
\eq{Phi:new} serves as the order parameter for the deconfinement
phase transition, if Mandelstam strings are condensed, {\it i.e.}
the coupling constant $\kappa$ should be larger than
the critical value $\kappa_c (B)$.

Now let us show that the expectation value of the old monopole
creation operator \eq{Phi} behaves as an order parameter for the
Dirac string with a fixed length. We generate the compact Abelian
Higgs model for $\gamma = 0.3$, $q=1$ and couplings $\beta=0.6$
(confinement phase) and $\beta=1.2$ (deconfinement phase).

\begin{figure}[!htb]
\vspace{2mm}
\begin{center}
\epsfxsize=6.0cm \epsffile{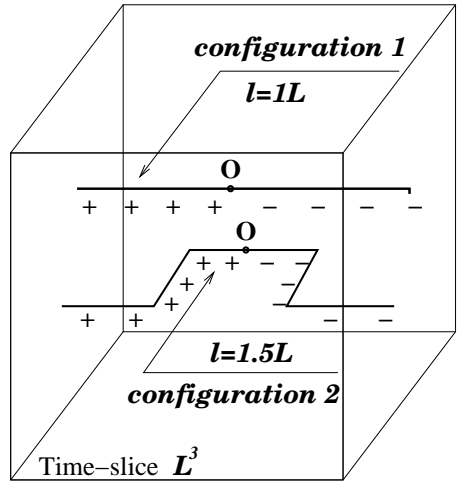}
\end{center}
\caption{Two configurations of the Dirac strings used in our
simulations on the lattice with anti--periodic boundary conditions.}
\label{fig:confs}
\end{figure}

\begin{figure}[!htb]
\vspace{2mm}
\begin{center}
\epsfxsize=8.cm \epsffile{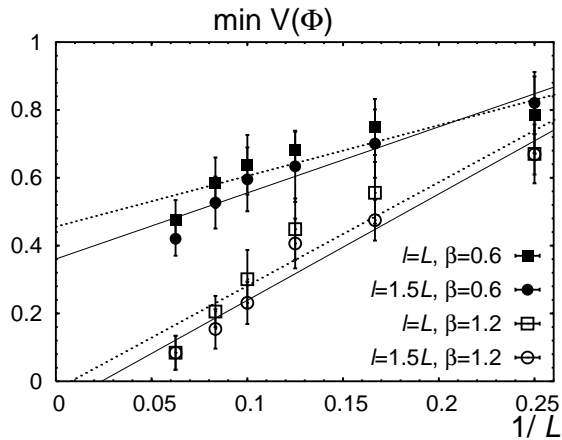}
\end{center}
\caption{The minimum of effective monopole potential~\eq{eff:potential}
as a function of the inverse lattice size $1/L$ for the Dirac strings with
various lengths $\ell$.}
\label{fig:minV}
\end{figure}

We use Dirac strings of the given length in proportion $1:1.5$ on
the $4^4,6^4,8^4,10^4,12^4,16^4$ lattices. The form of the Dirac
strings is shown in Figure~\ref{fig:confs}.
For each value of $\beta$ and for the fixed form of the Dirac string
we took average over 3000 values of the monopole creation operator.
The dependence of the minimum of the effective potential, $\min V( \Phi)$,
on lattice of the size $L$ for the fixed length of Dirac string $\ell$
is shown in Figure~\ref{fig:minV}.
We fitted the data for $\Phi_{\min}$ by the formula
$\Phi_{\min}=a L^b + \Phi_{\min}^{inf}$, where $a,b$ are
fitting parameters. It occurs that $b=-1$ within statistical errors.
The resulting values of $\Phi_{\min}^{inf}$ are collected in Table~1.
\begin{table}[htbp]
\begin{center}
\begin{tabular}{|c|c|c|} \hline
\hline $~~~\beta~~~$ & $\ell$  & $\Phi_{\min}^{\mathrm{inf}}$ \\
\hline $0.6$ & $L$   & $0.45(7)$ \\
\hline $0.6$ & $1.5L$& $0.36(5)$ \\
\hline $1.2$ & $L$   & $-0.03(6)$ \\
\hline $1.2$ & $1.5L$& $-0.07(5)$ \\
\hline
\end{tabular}
\end{center}
\caption[Table] {The minimum of the monopole potential,
$\Phi_{\min}^{inf}$, based on the old operator $vs.$ the gauge coupling $\beta$
and the Mandelstam string length $\ell$.}
\end{table}
We obtain that $\Phi_{\min}^{inf}$ vanishes at the point of the
phase transition. Thus the operator \eq{Phi:new} serves as the
order parameter for the deconfinement phase transition provided
the length of Dirac string is fixed.

\newpage

\section{Conclusions}

The new monopole creation operator proposed in Ref.~\cite{FrMa99}
can be used as a test of the monopole condensation in the theories
with electrically charged matter fields. Our calculations indicate
that the operator should be defined in the phase where the
Mandelstam strings are condensed. The minimum of the effective
potential corresponds to the value of the monopole field which is
zero in deconfinement phase and nonzero in the confinement phase.

In the infinite volume limit the potential corresponding to the
old version of the monopole creation operator shows the same
features as the potential calculated with the use of the new
operator. The shape of the old effective potential depends on the
length of the Dirac string. This fact indicates that the Dirac
strings with different shapes provide different monopole creation
operators and all of them can serve as order parameters for the
confinement--deconfinement phase transition.

\section*{Acknowledgments.}

The authors are grateful to F.V. Gubarev for useful discussions.
V.A.B. and M.I.P were partially supported by grants RFBR
02-02-17308, RFBR 01-02-17456, INTAS 00-00111 and CRDF award
RP1-2364-MO-02. M.N.Ch. is supported by JSPS Fellowship P01023.

\end{document}